\documentclass[aps,prl,twocolumn,superscriptaddress,longbibliography,footinbib]{revtex4-2}
\usepackage{graphicx}
\usepackage{dcolumn}
\usepackage{bm,color}
\usepackage{amssymb}
\usepackage{amsmath}
\usepackage{extarrows}
\usepackage[colorlinks=true,linkcolor=blue,urlcolor=blue,citecolor=blue]{hyperref}
\usepackage[normalem]{ulem} 

\begin{document}
\title{Edge-state interferometry as a probe of local flux in isolated quantum Hall systems}

\author{ Botao Wang}
\email[]{botao.wang@ulb.be}
\affiliation{Center for Nonlinear Phenomena and Complex Systems, Universit\'e Libre de Bruxelles, CP 231, Campus Plaine, 1050 Brussels, Belgium}
\affiliation{International Solvay Institutes, 1050 Brussels, Belgium}
\author{Nathan Goldman}
\email[]{nathan.goldman@lkb.ens.fr}
\affiliation{Center for Nonlinear Phenomena and Complex Systems, Universit\'e Libre de Bruxelles, CP 231, Campus Plaine, 1050 Brussels, Belgium}
\affiliation{International Solvay Institutes, 1050 Brussels, Belgium}
\affiliation{Laboratoire Kastler Brossel, Collège de France, CNRS, ENS-Universit\'e PSL, Sorbonne Université, 11 Place Marcelin Berthelot, 75005 Paris, France}
\author{Andr\'e Eckardt}
\email[]{eckardt@tu-berlin.de}
\affiliation{Technische Universit\"at Berlin, Institut f\"ur Theoretische Physik, Hardenbergstr.\ 36, 10623 Berlin, Germany}

\date{\today}

\begin{abstract}
	Quantum point contacts (QPCs) are essential tools for transport experiments in solid-state systems, enabling the detection of fractional charges and anyonic braiding statistics. Realizing analogous transport setups in isolated quantum-simulation platforms, such as ultracold atoms, remains challenging, since it typically requires coupling to external reservoirs. Here we show that the scattering properties of chiral edge states at a QPC can instead be extracted directly from the stationary edge currents of an isolated, reservoir-free lattice system. Exploiting the sensitivity of this scattering to Aharonov-Bohm-type phases, we propose an equilibrium protocol to detect local magnetic fluxes from ground-state edge currents. We further introduce a dynamical scheme, robust against finite temperature and particle-number fluctuations, based on the post-quench evolution following a sudden potential-bias removal. Since anyonic excitations are themselves associated with a local, quantized magnetic flux, our approach should extend to probing anyonic statistical phases in quantum-engineered platforms.
\end{abstract}

\maketitle

\textit{Introduction---}
Quantum point contacts (QPCs) -- narrow constrictions that act as a zero-dimensional defect in an otherwise extended conductor -- serve as fundamental tools for transport experiments in solid-state systems~\cite{2021Feldman,2021Carrega}. A QPC gives rise to quantized conductance even in the absence of a magnetic field~\cite{1988Wees,1988Wharam}, and, in topological systems hosting chiral or helical edge modes, can additionally induce backscattering between edge modes located on opposite edges of the system. This effect underlies electronic Fabry-P\'erot~\cite{1989Kivelson,1997Chamon,2006Kim,2011Halperin,2012Rosenow,1997Schuster,2008Deviatov,2009Ofek,2009Zhang_FP,2009McClure,2012McClure,2015Choi,2019Nakamura} and Mach--Zehnder~\cite{2003Ji,2006Neder,2007Neder,2007Neder_dephasing,2007Roulleau,2008Roulleau,2008Roulleau_coherence,2008Bonderson} interferometers, which, together with other approaches~\cite{2020Bartolomei,2023Ruelle,2023Lee,2023Jonckheere,2024Werkmeister}, have enabled the observation of fractional charges~\cite{1997Picciotto,1997Saminadayar} and anyon braiding phases in fractional quantum Hall states~\cite{2020Nakamura,2023Nakamura,2023Willett,2023Kundu,2021Feldman,2021Carrega}. Despite these breakthroughs in electronic systems, realizing analogous phenomena in isolated quantum-simulation platforms, such as ultracold atomic gases, remains a major challenge.

Recently, transport experiments with ultracold atoms have achieved considerable progress, particularly in two-terminal configurations modeling a single QPC~\cite{2015Chien,2017Krinner,2022Amico}. 
Connecting two cold atom reservoirs has enabled the realization of thermoelectric heat engines~\cite{2013Brantut}, quantized conductance with and without dissipation~\cite{2015Krinner,2019Lebrat,2019Corman,2023Huang}, as well as the transport of spin~\cite{2020Sekino}, entropy~\cite{2024Fabritius,2024Mohan} and dark-state transport~\cite{2024Talebi}, to name a few. Complementary theoretical studies have explored transport through Hofstadter~\cite{2019Salerno} and sawtooth lattices \cite{2021Pyykkonen} coupled to external reservoirs. 
However, simultaneously implementing a multi-terminal transport setup and engineering artificial gauge fields to access topological states with chiral or helical edge modes poses severe experimental challenges.
For instance, while time-periodic driving (e.g.\ Floquet engineering) is typically required to generate artificial gauge fields~\cite{2011Dalibard,2014Goldman,2016Goldman,2017Eckardt,2017Zhang_Manipulating, 2018Aidelsburger, 2019Zhang_rev,2019Cooper, 2019Ozawa,2021Weitenberg}), it is often challenged by unwanted non-equilibrium processes when the system is coupled to reservoirs~\cite{2009Hone}.

\begin{figure}
	\centering\includegraphics[width=1\linewidth]{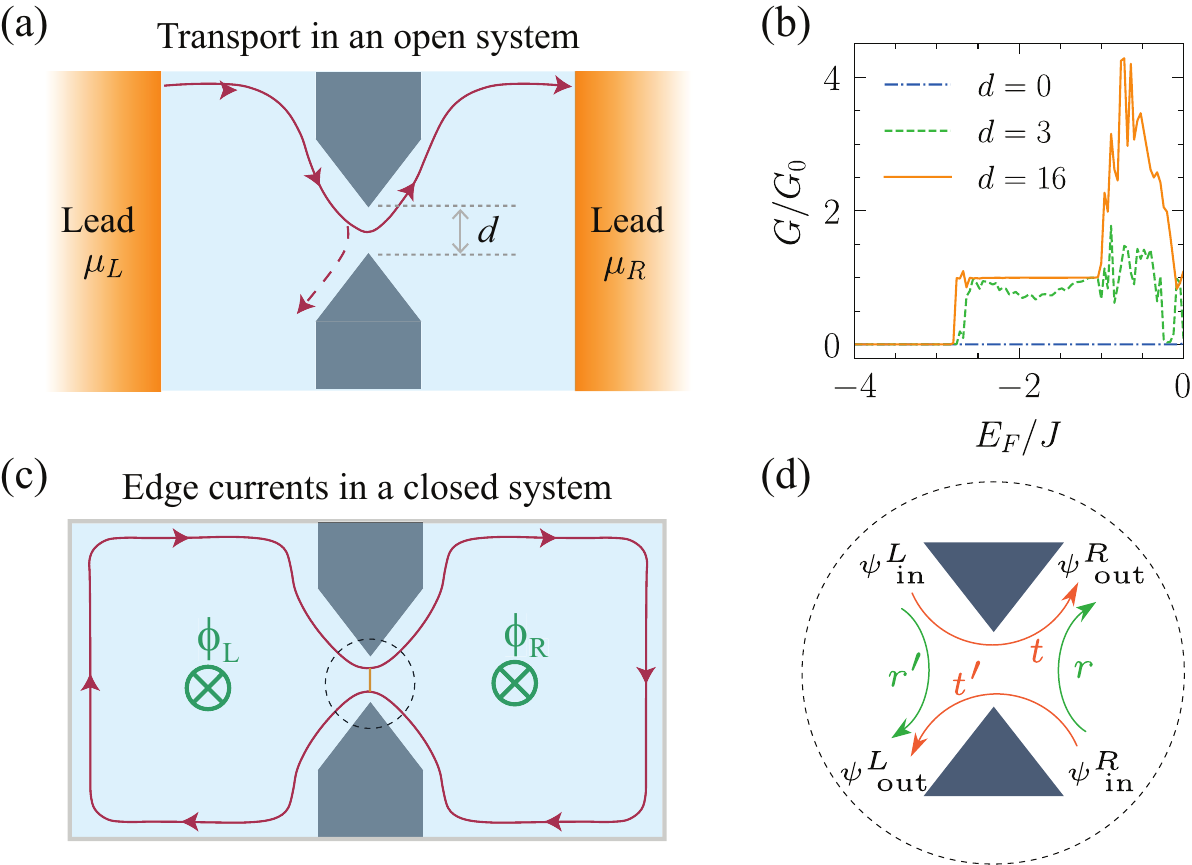} 
	\caption{(a) Schematic of a QPC system coupled to two leads with chemical potentials $\mu_L$ and $\mu_R$.  (b) Conductance (in units of the conductance quantum $G_0=e^2/h$) as a function of Fermi energy $E_F$ for different QPC distances $d$. These results are obtained from simulating a $25\times 16$ Hofstadter lattice at quarter flux $\alpha=1/4$ attached to two semi-infinite leads, using the Kwant software package~\cite{Groth_2014}. (c) Schematic of a closed system hosting chiral edge modes. (d) Sketch of the scattering problem near the QPC. Solving this scattering problem in terms of the stationary edge currents yields the transmission rate of the QPC, and enables the determination of the flux difference $\delta\phi=\phi_R-\phi_L$. } 
	\label{fig_sketch}
\end{figure}

In this work, we show that it is possible to investigate the scattering properties of chiral edge states at a QPC directly within isolated lattice systems, where no reservoir-driven particle flow needs to be engineered. Specifically, the scattering properties of chiral edge modes at a QPC are found to influence the ground state properties of the system, making it experimentally accessible through measurements of stationary currents. Exploiting the fact that these scattering properties are influenced in a well-defined fashion by an Aharonov-Bohm-type phase, we propose an equilibrium protocol to measure local magnetic fluxes [see Fig.~\ref{fig_sketch}(c)] via stationary edge currents. This method relies on comparing an unknown flux to a tunable reference flux, in much the same way that an unknown mass is weighed against known reference weights on a balance scale. Moreover, we develop a second, dynamic scheme that monitors the post-quench dynamics following the sudden removal of an initial potential bias in an initial equilibrium state. Unlike the first approach, this dynamic probe remains robust against both finite temperatures and particle number fluctuations. Our framework offers a realistic way for measuring Aharonov-Bohm-type fluxes and opens new possibilities for probing non-trivial anyonic statistical phases.

\textit{Model---}We consider a tight-binding square lattice threaded by a homogeneous magnetic flux $\phi\!=\!2\pi\alpha$ per plaquette, supplemented by tunable, inhomogeneous local fluxes. The system is described by the Harper-Hofstadter Hamiltonian~\cite{1955Harper,1976Hofstadter}
\begin{align}
	\hat{H}= & -J\sum_{m,n}\left(e^{-i\theta_{m,n}}\hat{a}_{m+1,n}^{\dagger}\hat{a}_{m,n}+\hat{a}_{m,n+1}^{\dagger}\hat{a}_{m,n}+\text{h.c.}\right)
	\nonumber \\ &
	+ \sum_{m,n} V_{m,n}\hat{n}_{m,n}, 
	\label{H}
\end{align}
where $\hat{a}_{m,n}^{\dagger}, \hat{a}_{m,n}$ and $\hat{n}_{m,n}=\hat{a}_{m,n}^{\dagger}\hat{a}_{m,n}$ are the creation, annihilation, and number operator for fermions on site $(m,n)$, with integer $x$ and $y$ coordinates $m=0,1,\ldots, L_x-1$ and $n=0,1,\ldots, L_y-1$, respectively. We define the tunneling parameter $J$, the on-site potential $V_{m,n}$, as well as the Peierls phases $\theta_{m,n}$ for tunneling from site $(m,n)$ in the positive $x$ direction. 
Choosing $\theta_{m,n}=\phi n$ leads to the background magnetic flux $\phi=2\pi\alpha$ per plaquette. Moreover, the flux of a given plaquette can be further modified by applying additional Peierls phases~\cite{2018Wang,2018Raviunas,2022Wang}. 
In the following we assume tunable additional fluxes $\phi_L$ and $\phi_R$ centered within the left and right regions, respectively. The model (\ref{H}) can be realized using Floquet engineering~\cite{2018Wang}. In the absence of interactions, which is natural for spin-polarized ultracold atomic gases, Floquet heating does not pose a severe challenge~\cite{2017Eckardt}.

A QPC geometry can be implemented by properly designing large positive on-site potentials $V_{m,n}$ in specific regions (in our simulations $V_{m,n}\!=\!100J$), as depicted by the gray area in Figs.~\ref{fig_sketch}(a) and (c). The distance of the barriers is denoted by $d$ in unit of lattice constant. In cold atom experiments, such QPC potentials can be readily realized, for example, by using digital mirror devices or spatial light modulators~\cite{2021Navon}. In the following, we set $\hbar=1$, and use the lattice constant $a$ and the tunneling parameter $J$ as the units of length and energy, respectively.

As a paradigmatic model of a Chern insulator~\cite{2019Cooper}, the Hofstadter Hamiltonian~(\ref{H}) exhibits Chern bands separated by bulk energy gaps, giving rise to an incompressible state when individual bands are filled completely with non-interacting fermions. Under open boundary conditions, however, the edges of the system remain compressible due to the presence of chiral edge modes connecting adjacent bulk bands. The resulting chiral edge currents can be clearly revealed by considering the following local current operators
\begin{align}
	\hat{j}_{m,n}^{x} &= iJ_x\left(e^{-i\theta_{m,n}}\hat{a}_{m,n}^{\dagger}\hat{a}_{m+1,n}-\text{h.c.}\right),
	\\
	\hat{j}_{m,n}^{y} & = iJ_{y}\left(\hat{a}_{m,n}^{\dagger}\hat{a}_{m,n+1}-\text{h.c.}\right),
\end{align}
which represent the currents in the $x$- and $y$-directions, respectively. These expressions for the local currents can be derived from the time derivative of the particle density $n_{m,n}$ \cite{1991Silva} and have been measured in optical lattices~\cite{2014Atala,2023Impertro,2025Impertro}. We denote their expectation value by $j_{m,n}^{\eta} =\langle \hat{j}_{m,n}^{\eta} \rangle$ with $\eta=x,y$. Before turning to the effect of the QPC on these stationary currents, we note that studying currents rather than densities offers a further advantage:~since only chiral edge modes carry steady-state currents, currents act as a natural filter that isolates their contribution.  

\textit{Transport in an open system---}A typical transport setup often requires the coupling to external reservoirs. As illustrated in Fig.~\ref{fig_sketch}(a), we consider a Hofstadter lattice with uniform flux $\alpha=1/4$ connected to two infinite leads as an example. The transport properties are commonly analyzed within the Landauer-B\"{u}ttiker formalism~\cite{1957Landauer,1985Buttiker,1999Imry}. In the absence of a QPC, a quantized Hall conductance can be obtained using the recursive Green's function method~\cite{Datta_2005,2019Salerno}, or through numerical simulations based on the Kwant package~\cite{Groth_2014}. Using the latter method, we consider the case where the Fermi energies of both reservoirs lie within the bulk gap separating the lowest band, with Chern number $C\!=\!1$, from the excited bands. In this regime, a single edge mode provides the only transport channel, yielding a quantized conductance [orange solid line in Fig.~\ref{fig_sketch}(b)]. In contrast, when the Fermi energy is set within a bulk band, multiple transmitting channels contribute and thus give rise to metallic behavior accompanied by non-quantized conductance~\cite{2019Salerno}. In the presence of a QPC, however, we observe a reduced conductance; see the green dashed line in Fig.~\ref{fig_sketch}(b). This can be attributed to backscattering at the QPC. We show below that, instead of considering such an open setup, the conductance can still be inferred from stationary currents in an isolated system.

\begin{figure}
	\centering\includegraphics[width=1\linewidth]{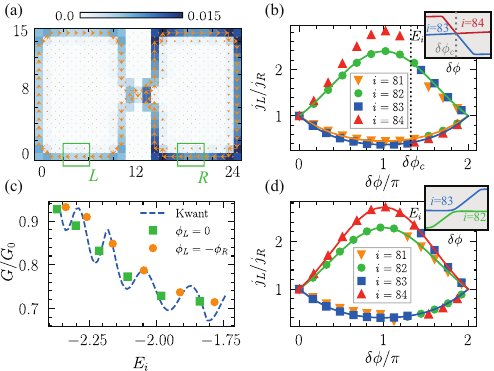} %
	\caption{(a) Density and current distribution of eigenstate $i=83$ at $\phi_R=\pi/2, \phi_L=0$ for a Hofstadter lattice of $25\times 16$ sites at quarter flux $\alpha=1/4$. The lower boxes $L$ and $R$ define the region for the calculation of the currents $j_{\mathcal{R}}=\sum_{m,n\in\mathcal{R}}j_{m,n}^{x}/N_{C}$ with $\mathcal{R}=L,R$ and $N_C$ the number of columns within the boxes. (b) Ratio of stationary currents as a function of $\delta\phi$ for $(\phi_L, \phi_R)=(0, \delta\phi)$. Discrete data correspond to currents of different edge modes $i$ obtained from numerical simulations, and the dashed lines represent the analytic results with different fitting parameter $r$ [Eq.~(\ref{gamma_result})]. Inset: $i$-th eigenvalue as a function of flux difference $\delta\phi$; see Fig.~\ref{fig_supp} for the full spectrum. The vertical dotted line marks the avoided crossing at $\delta\phi_c\simeq1.32\pi$ for $i=83$, which is responsible for the abrupt jump in the blue squares in (b). The avoided crossing points are slightly shifted for different state indices, and $\delta\phi_c/\pi\simeq1.42$ and $1.38$ for $i=81$ and $i=82$, respectively.  (c) Transmission probability as a function of the eigenenergies $E_i$. (d) Same as (b) but for $(\phi_L, \phi_R)=(-\delta\phi/2, \delta\phi/2)$. The different configurations in (b) and (d) give rise to the same flux difference $\delta \phi$.  }
	\label{fig_edge}
\end{figure}

\textit{Edge-mode scattering at a QPC---} As depicted in Figs.~\ref{fig_sketch}(c) and (d), we now focus on a closed system where a QPC connects chiral edge modes on the opposite side. There is an incoming (``in'') and and outgoing (``out'') mode on the left (L) and the right (R) of the QPC, whose amplitudes form the elements of two vectors, $\ensuremath{\mathbf{\Psi}}_{\text{in}}=(\psi_{\text{in}}^{R},\psi_{\text{in}}^{L})^{T}$ and $\ensuremath{\mathbf{\Psi}}_{\text{out}}=(\psi_{\text{out}}^{R},\psi_{\text{out}}^{L})^{T}$. They are related by the unitary scattering matrix $\mathbf{S}$,
\begin{equation}
	\ensuremath{\mathbf{\Psi}}_{\text{out}}=\mathbf{S}\ensuremath{\mathbf{\Psi}}_{\text{in}}, ~~~~\mathbf{S}=\left(\begin{array}{cc}
		r & t\\
		t & r
	\end{array}\right).
	\label{S}
\end{equation}
Here the complex numbers $r=|r|e^{i\theta_{r}}$ and $t=|t|e^{i\theta_{t}}$ describe reflection and transmission, respectively, which are assumed to be symmetric with respect to both sides $L$ and $R$. Owing to the unitarity of $\mathbf{S}$, it follows straightforwardly that the phases satisfy $\cos(\theta_{t}-\theta_{r})=0$; see End Matter.

Since on each side the outgoing and incoming edge modes correspond to two ends of a loop, the associated wave function have equal magnitudes and therefore differ only by a phase factor, i.e.\ $\psi_{\text{out}}^{R/L}=e^{i\varphi_{R/L}}\psi_{\text{in}}^{R/L}$. Each phase consists of a dynamical and a geometric contribution, and the latter includes the local magnetic flux $\phi_{L/R}$ threading the corresponding edge-mode loop. 
Assuming that only the geometric phases differ between the two sides, as is the case when a localized flux is introduced at the center of one (or both) loops, one has $\varphi_R-\varphi_L=\phi_R-\phi_L\equiv \delta \phi$. Using the unitarity of $\mathbf{S}$, we obtain
\begin{equation}
	\frac{\psi_{\text{in}}^{L}}{\psi_{\text{in}}^{R}}=\pm i\frac{(1-e^{i\delta\phi})|r|\pm\sqrt{(e^{i\delta\phi}+1)^{2}|r|^{2}-4e^{i\delta\phi}}}{2\sqrt{1-|r|^{2}}e^{i\delta\phi}},
	\label{gamma_result}
\end{equation}
where $|r|^2$ and $|t|^2=1-|r|^2$ denote the reflection and transmission probabilities, respectively. The above equation describes a generic scattering problem of chiral edge modes encountering at a QPC. 
The stationary edge currents on both sides, $L$ and $R$, of the system, are directly proportional to $\psi_\text{in}^L$ and $\psi_\text{in}^R$, respectively (as states occupying bulk modes do not contribute to the edge current). Measuring these currents therefore directly yields the ratio $\psi_\text{in}^L / \psi_\text{in}^R$ as a function of the flux difference $\delta\phi$, from which one can extract the reflection probability $|r|^2$, and hence the transmission probability $|t|^2=1-|r|^2$.
Conversely, once $|r|^2$ is determined (see below), the magnetic flux difference can be extracted from stationary-current measurements via Eq.(\ref{gamma_result}).

\textit{Stationary currents of a single edge mode---}
As a concrete application, we investigate the fate of a single edge mode in a Hofstadter lattice in the presence of a QPC.
Considering a $25\!\times\!16$ system with $\phi_L\!=\!0$ and $\phi_R\!=\!\delta\phi\!=\!\pi/2$, we plot the density and current distributions of a representative single edge mode (state index $i=83$, counted from the bottom of the spectrum) in Fig.~\ref{fig_edge}(a). The QPC geometry, together with a finite flux difference $\delta\phi\neq0$, induces an asymmetry in both the current and density distributions. Such an asymmetric behavior alternates between edge states with even and odd indices. 
To quantify the dependence of the currents on the flux $\delta\phi=\phi_R$, we plot the ratio of the current between the left and the right $j_L/j_R=\left.\sum_{(m,n)\in L}j_{m,n}^{x}\right/\sum_{(m,n)\in R}j_{m,n}^{x}$ as a function of $\delta\phi$ in Fig.~\ref{fig_edge}(b). 
We observe an abrupt jump in the current ratio at $\delta\phi_c$, after which the ratio takes values larger (smaller) than 1 for odd (even) state indices. This behavior is attributed to an avoided crossing in the spectrum (see inset of Fig.~\ref{fig_edge}(b) and End Matter). The avoided crossing is associated with the quantized charge pumping induced by the insertion of a Laughlin-type local flux in the lattice~\cite{2018Wang,2018Raviunas,2022Wang}. As a consequence, to compensate for this pumping, the properties of edge mode $i$ are taken over by mode $i+1$ after passing through the avoided crossing. 

Fitting the resulting continuous current ratio $j_L/j_R=|\psi_{\text{in}}^{L}/\psi_{\text{in}}^{R}|^2$ to the analytical expression~(\ref{gamma_result}), with the transmission coefficient $|r|$ as the only fitting parameter, we find excellent agreement [see Fig.~\ref{fig_edge}(b)]. Namely, the fitting curve (blue solid line) continuously connects the current ratios corresponding to $i=83$ for $\delta\phi<\delta\phi_c$ and $i=84$ for $\delta\phi>\delta\phi_c$.
Moreover, the extracted value of $1-|r|^2$ agrees well with the conductance obtained from an independent transport simulation of the QPC in a system coupled to infinite leads~\cite{2014Groth,2019Salerno}, as shown in Fig.~\ref{fig_edge}(c). 
The same agreement holds for a modified flux configuration with $\phi_R=-\phi_L=\delta\phi/2$ [see Figs.~\ref{fig_edge}(c, d)].
This consistency confirms that, in a closed system, the ground-state edge currents are determined by the scattering properties of the QPC together with the flux imbalance $\delta \phi$, which constitutes a first key result of this work.

We now exploit this correspondence to propose two complementary interferometric schemes for measuring magnetic flux differences in isolated systems. These schemes offer an alternative to existing solid-state approaches, which rely on lead-induced transport currents~\cite{2021Feldman,2021Carrega} and which are not straightforward to implement in quantum-simulation platforms.

\textit{Chern-insulator ground states---}
While populating a single edge state is feasible in certain systems, such as small photonic devices with high frequency resolution~\cite{2019Ozawa_photon,2013Hafezi_edge}, it remains significantly more challenging in generic quantum-matter experiments, such as cold atoms in optical lattices. We therefore consider a more realistic setting based on Chern-insulator ground states, with the Fermi energy placed within the lowest bulk gap such that the entire lowest band, together with a subset of edge states, is filled. In this configuration, the edge current is given by the sum of the currents of the edge states below the Fermi energy. This approach relies on two assumptions: (i) precise control of the particle number; and (ii) a temperature below the finite-size level spacing between neighboring edge states. We will relax these assumptions below when introducing a second, dynamical scheme for measuring flux differences.

\begin{figure}
	\centering\includegraphics[width=1\linewidth]{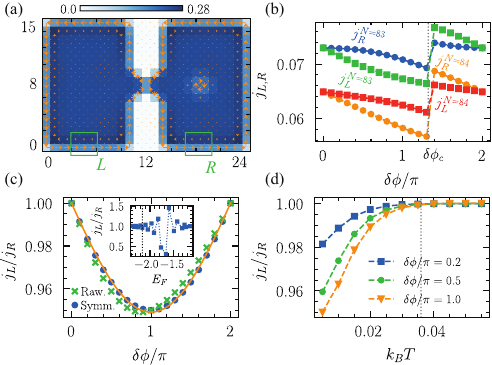}
	\caption{
    (a) Spatial density and current distributions of the Chern insulator ground state with $N=83$ particles at $\delta\phi=\phi_R=\pi/2$. The lower boxes $L$ and $R$ indicate the regions where the left and right currents $j_L$ and $j_R$ are evaluated, respectively. (b) Left and right currents $j_{L,R}$ as a function of flux imbalance $\delta\phi$ for two particle numbers $N=83$ and $N=84$. The discontinuities are associated with an avoided crossing in the spectrum (see Fig.~\ref{fig_edge}(b) and \ref{fig_supp}), which can be compensated for by connecting solutions for $N=83$ and $N=84$ particle sectors. The vertical dotted line denotes the avoided crossing point at $\delta\phi\simeq1.32\pi$. (c) Ratio of the stationary currents $j_L/j_R$ as a function of $\delta\phi$. Green crosses correspond to results obtained using $N=83$ for $\delta\phi<\delta\phi_c$ and $N=84$ for $\delta\phi>\delta\phi_c$. Blue dots represent the symmetrized results obtained by considering an inverted flux insertion $-\delta\phi$ to compensate for the charge pumping effect. The solid line is the linear fit based on Eq.~(\ref{gamma_result}) which leads to $|r|=0.026$. Inset: Current ratio $j_L/j_R$ at $\delta\phi=\pi/2$ as a function of the Fermi energy $E_F$. The vertical dotted line indicates the Fermi energy corresponding to $N=83$, which is used in the main panel. (d) Current ratio as a function of temperature $k_BT$ for different values of $\delta\phi$. These curves collapse onto each other once the temperature becomes larger than the energy level splitting $E_{83}-E_{82}\simeq0.036$ indicated by the vertical dotted line. Here we consider non-interacting fermions in a $25\times 16$ lattice with $\alpha=1/4$ and $\phi_L=0,\phi_R=\delta\phi$.} 
	\label{fig_gs}
\end{figure}

A typical spatial density and current distribution of a Chern insulator ground state at $\delta\phi=\phi_R=\pi/2$ is depicted in Fig.~\ref{fig_gs}(a). The total edge currents, obtained by summing over all occupied states, are plotted in Fig.~\ref{fig_gs}(b). Similarly, an abrupt jump appears at the avoided crossing point $\delta\phi=\delta\phi_c$. Such discontinuities can be removed by connecting results measured for different (neighboring) particle numbers, so that a smooth curve is obtained.
The resulting continuous curves give rise to the current ratio $j_L/j_R$ shown by the green crosses in Fig.~\ref{fig_gs}(c), which is asymmetric with respect to $\delta\phi$. This asymmetry is attributed to the charge pumping effect induced by varying the flux $\delta\phi$, and can be compensated for by considering symmetrized currents $j_\ell^{\rm symm}=[j_\ell(\delta\phi)+j_\ell(-\delta\phi)]/2$ with $\ell=\{L,R\}$, as indicated by the blue dots in Fig.~\ref{fig_gs}(c).  In this case, it shows qualitatively similar behavior to that of individual edge states [Fig.~\ref{fig_edge}(b)], allowing for the determination of the flux difference. Even though the total signal shown here is small due to the alternating sign of the edge current, it can be largely enhanced by fine-tuning the Fermi energy, as shown in the inset of Fig.~\ref{fig_gs}(c). Specifically, the signal becomes stronger when the Fermi energy is closer to the middle of the energy gap. We note that for a Chern insulator ground state, summing over the currents of alternating edge states introduces sensitivity to finite temperatures. As shown in Fig.~\ref{fig_gs}(d), the dependence of the current ratio on $\delta\phi$ is suppressed once the temperature exceeds the small energy level splitting.

\textit{Robust scheme based on post-quench dynamics---}
It is desirable to find a signature of the $\delta\phi$-induced imbalance of the edge states that can be measured under realistic experimental conditions, i.e.\ without resolving them individually at near-zero temperatures and precise control of the particle number. For that purpose, we propose the following dynamical protocol. Initially, the system is prepared in its equilibrium state in the presence of an additional energy offset between left and right, given by the additional on-site energy $V_L$ on all lattice sites on the left-hand-side of the system. For $V_L$ small compared to the band gap, this offset primarily affects the gapless edge modes. Provided $V_L$ is also large compared to the small finite-size level splitting of the edge modes, it causes an increased population of sites near the edge in the left (right) half of the system for $V_L<0$ ($V_L>0$). In a second step, $V_L$ is then suddenly switched off and the current imbalance between left and right is recorded as a function of time. Note that while it is also possible to record the density imbalance, the current imbalance has the advantage that it is less influenced by residual contributions from bulk states. When the individual edge modes are imbalanced, as is the case for $\delta\phi\ne 0$ modulo $2\pi$, quenching $V_L$ to zero will predominantly populate those edge states with a larger weight on the left (right) of the system. 
As a result, such an imbalance persists in the time-averaged current and in the long-time limit. 
The more strongly the individual edge modes are localized on one side of the system, the stronger this signal becomes. In turn, for $\delta\phi=0$ the individual edge modes are symmetric, causing the current imbalance to average out over time. 

\begin{figure}
	\centering\includegraphics[width=1\linewidth]{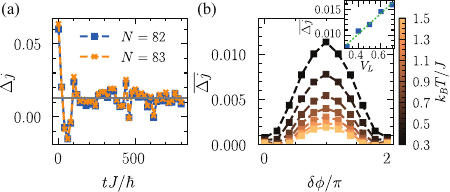}
	\caption{(a) Current difference $\Delta j = j_R-j_L$ as a function of evolution time $t$ for two particle numbers $N=82$ and $N=83$. Here we consider $k_BT=0.1, \phi_R=\pi,\phi_L=0$ and initial bias potentials $V_L=-0.5$ as an example. The currents $j_L$ and $j_R$ are evaluated by summing over all occupied states governed by the Fermi-Dirac distribution at finite temperature $k_B T$. The gray solid line represents the time-averaged current difference $\overline{\Delta j}$ over the entire evolution period up to $t=800$. (b) Time-averaged current difference $\overline{\Delta j}$ as a function of $\delta\phi$ for various temperatures. Inset: $\overline{\Delta j}$ versus bias potentials $V_L$ at $k_BT=0.2, \phi_R=0.8\pi$ and $\phi_L=0$. We find a linear increase of current difference as a function of $V_L$ within the range $\delta<V_{L}<\Delta_{g}$, where $\delta$ and $\Delta_g$ denotes the edge mode splitting and bulk gap, respectively. } 
	\label{fig_dyn}
\end{figure}

Fig.~\ref{fig_dyn}(a) shows the time evolution of the current imbalance after the potential offset of $V_L=-0.5$ is switched off, with the initial state being prepared at a finite temperature $k_BT=0.1$ and a phase difference of $\delta\phi=\pi$. We can clearly see that the imbalance fluctuates around a finite mean value. This value does not depend on the precise particle number, as can be seen from the comparison of the results for $N=82$ and $83$ particles. The time-averaged current imbalance is directly connected to the flux imbalance and remains visible also for temperatures of the order of the tunneling energy [Figs.~\ref{fig_dyn}(b)]. Its dependence on the offset $V_L$ is roughly linear (see inset).

\textit{Conclusions---}We have demonstrated that the scattering properties of chiral edge states at a QPC can be directly inferred from the ground-state properties of an isolated lattice system, bypassing the experimental challenge of engineering external particle reservoirs. By exploiting the sensitivity of these edge states to Aharonov-Bohm-type phases, we proposed an equilibrium protocol to measure local magnetic fluxes via stationary current distributions. To overcome the constraints of finite temperatures and precise particle-number control, we further designed a robust dynamical scheme based on the post-quench dynamics following a sudden potential-bias removal. Beyond flux probing, since anyonic excitations are themselves naturally associated with a local, quantized magnetic flux, our method should extend directly to their detection, opening a promising pathway toward measuring fractional charges and anyonic braiding phases in both synthetic quantum matter and solid-state systems~\cite{2021Feldman,2021Carrega,2023Schiller,2026Henzinger,2026Kim}.

\paragraph{Acknowledgments.}
We thank Iacopo Carusotto, Cecile Repellin, Meng-Zi Huang and Ignacio Cirac for helpful discussions. B.W. and N.G. acknowledge support by the FRS-FNRS (Belgium), the ERC Starting Grants TopoCold and LATIS, the EOS project CHEQS, the Fondation ULB and the ANR PEPR
Grant QUTISYM ANR-23-PETQ-0002. A.E. acknowledge support from the Deutsche Forschungsgemeinschaft (DFG) via the Research Unit FOR 5688  under Project No. 521530974. 

\bibliography{mybib}


\setcounter{equation}{0}
\setcounter{figure}{0}
\renewcommand{\theequation}{A\arabic{equation}}
\renewcommand{\thefigure}{A\arabic{figure}}
\renewcommand{\thesection}{A\arabic{section}}
\renewcommand{\thesubsection}{\thesection.\arabic{subsection}}
\renewcommand{\thesubsubsection}{\thesubsection.\arabic{subsubsection}}

\begin{center}
	{\bfseries End Matter}
\end{center}
\setcounter{section}{0} 
\renewcommand{\thesection}{\Alph{section}} 

\refstepcounter{section}

\section*{A. Scattering at a QPC}
\label{Appd_A}

Consider a generic setup with a QPC at the center of the system, as depicted in Fig.~\ref{fig_sketch}(d), the incoming and outgoing wave functions, denoted by $\ensuremath{\mathbf{\Psi}}_{\text{in}}=(\psi_{\text{in}}^{R},\psi_{\text{in}}^{L})^{T}$ and $\ensuremath{\mathbf{\Psi}}_{\text{out}}=(\psi_{\text{out}}^{R},\psi_{\text{out}}^{L})^{T}$ respectively, are related by a scattering matrix $\mathbf{S}$ as defined in Eq.~(\ref{S}).

Probability conservation requires that $\ensuremath{\mathbf{\Psi}}_{\text{out}}^{\dagger}\ensuremath{\mathbf{\Psi}}_{\text{out}}=\ensuremath{\mathbf{\Psi}}_{\text{in}}^{\dagger}\ensuremath{\mathbf{\Psi}}_{\text{in}}$, which implies the unitarity of the scattering matrix, i.e. $\mathbf{S}^{\dagger}\mathbf{S}=\mathbf{S}\mathbf{S}^{\dagger}=1$. This imposes the following conditions for the reflection and transmission rate:
\begin{align}
	\begin{cases}
		|r|^{2}+|t|^{2}=1, \\ 
            |r'|^{2}+|t'|^{2}=1,\\
		r^{*}t'+t^{*}r'=0,  \\ 
		rt^{*}+t'r'^{*}=0.  \\ 
	\end{cases}
	\label{condition}
\end{align}

The incoming and outgoing wave function components are connected to each other and, thus, differ only by a phase factor, 
\begin{equation}
	\begin{cases}
		\psi_{\text{out}}^{R}= & e^{i\varphi_{R}}\psi_{\text{in}}^{R},\\
		\psi_{\text{out}}^{L}= & e^{i\varphi_{L}}\psi_{\text{in}}^{L}.
	\end{cases}
	\label{phi_out}
\end{equation}
The phase difference $\varphi_L-\varphi_R$ can be tuned by the Aharonov-Bohm type phases $\phi_L$ and $\phi_R$, given by the magnetic flux encircled by the loops formed by the left and right edge mode, respectively. For a configuration that is symmetric with respect to the left and right side, we have $r=r'$ and $t=t'$ and $\varphi_R-\varphi_L=\delta\phi=\phi_R-\phi_L$. Combining Eqs.(\ref{condition}) and (\ref{phi_out}), we then obtain
\begin{align}
	e^{i\theta_{R}}\psi_{\text{in}}^{R}= & r\psi_{\text{in}}^{R}+t\psi_{\text{in}}^{L}, \label{phi_R} \\
	e^{i\theta_{L}}\psi_{\text{in}}^{L}= & t\psi_{\text{in}}^{R}+r\psi_{\text{in}}^{L}. \label{phi_L}.
\end{align}
Using Eq.~(\ref{condition}) and defining
\begin{equation}
	r=|r|e^{i\theta_{r}}, ~~~~ t=|t|e^{i\theta_{t}},
	\label{rt}
\end{equation}
it follows that
\begin{equation}
	\cos(\theta_{t}-\theta_{r})=0.
	\label{rt_theta}
\end{equation}

Dividing Eq.~(\ref{phi_L}) by Eq.~(\ref{phi_R}) and using Eqs.~(\ref{rt}) and (\ref{rt_theta}), we obtain
\begin{align}
	\gamma e^{i\delta\phi}& =\frac{\psi_{\text{in}}^{R}|t|e^{i\theta_{t}}+\psi_{\text{in}}^{L}r|e^{i\theta_{r}}}{\psi_{\text{in}}^{R}r|e^{i\theta_{r}}+\psi_{\text{in}}^{L}|t|e^{i\theta_{t}}} =\frac{\pm i|t|+|r|\gamma}{|r|\pm i|t|\gamma},
	\label{gamma}
\end{align}
where we have introduced the notation $\gamma\equiv\psi_{\text{in}}^{L}/\psi_{\text{in}}^{R}$. Solving the above equation leads to the following expression of $\gamma$,
\begin{equation}
	\gamma=\pm i\frac{(1-e^{i\delta\phi})|r|\pm\sqrt{(e^{i\delta\phi}+1)^{2}|r|^{2}-4e^{i\delta\phi}}}{2\sqrt{1-|r|^{2}}e^{i\delta\phi}},
	\label{gamma_res}
\end{equation}
where $|r|^2$ and $1-|t|^2$ represents the reflection and transmission probability, respectively. By assigning the current ratio as $j_L/j_R=|\gamma|^2$, we find excellent agreement between the above analytic formula and our numerical simulations [see Figs.~\ref{fig_edge}(b,d)]. Note that the plus and minus sign in the numerator of Eq.~(\ref{gamma_res}) corresponds to the case of $j_L/j_R>1$ and $j_L/j_R<1$, respectively.

\section*{B. Energy spectrum}
\begin{figure}
	\centering\includegraphics[width=1\linewidth]{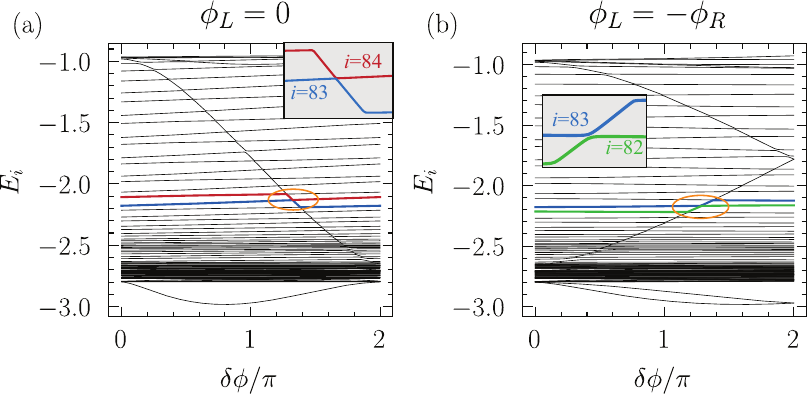} %
	\caption{ Energy spectrum near the lowest band as a function of flux difference $\delta\phi$ in the case of (a) $(\phi_L, \phi_R)=(0, \delta\phi)$ and (b) $(\phi_L, \phi_R)=(-\delta\phi/2, \delta\phi/2)$. Inset: zoom-in of a typical avoided crossing indicated by ellipse. Here we consider a $25\times 16$ Hofstadter lattice at quarter flux $\alpha=1/4$.}
	\label{fig_supp}
\end{figure}

Applying different configurations of additional flux, e.g., $(\phi_L, \phi_R) = (0, \delta\phi)$ or $(-\delta\phi/2, \delta\phi/2)$, leads to distinct spectral flow as a function of $\delta\phi$ (see Fig.~\ref{fig_supp}), which underlies the tailored quantized charge pumping~\cite{2018Wang}. The corresponding avoided crossings give rise to discontinuities in the steady-state currents (see, e.g., Fig.~\ref{fig_gs}(b)). These discontinuities are removed by connecting results for neighboring particle numbers, yielding smooth curves, as shown in Figs.~\ref{fig_edge}(b,d) and~\ref{fig_gs}(c).

\end{document}